\font\bfdue=cmbx10 scaled\magstep2
\font\bfu=cmbx10 scaled\magstep1

\font\bfz=cmbx10 scaled\magstep0

\font\tft=cmr10 scaled\magstep3
\font\tfdue=cmr10 scaled\magstep2
\font\tfu=cmr10 scaled\magstep1

\font\tfz=cmr10 scaled\magstep0

\font\ifdue=cmti10 scaled\magstep2
\font\ifu=cmti10 scaled\magstep1


\font\it=cmti10 scaled\magstep1

%

\def\frac#1#2{{$#1\over#2$}}

\def\der2parz#1#2{{\frac{\partial^2 #1}{\partial #2^2}}}


\def\mab{main\ asteroid\ belt\ }

\def\ast{asteroids\ }


%
\def\ez{\`e}


%
\def\es{\'e}

\def\is{\' \i}

\def\n{\~n}
\tolerance=2000 \hsize 5.8 truein

\vsize 8.9 truein

\baselineskip 21pt

\centerline {\tft Multifractal Fits to the Observed}

\centerline {\tft Main Belt Asteroid Distribution}

\vskip 1truecm

\centerline{\tfdue Adriano Campo Bagatin$^{1,2}$}

\centerline{\tfdue Vicent J. Mart\is nez$^3$}

\centerline{\tfdue Silvestre Paredes$^4$}

\vskip 1truecm

\centerline
{\tfu 1. Departamento de F\is sica, Ingenier\is a de Sistemas y Teor\is a de la Se\n al,}

\centerline
{\tfu E.P.S.A., Universidad de Alicante}

\centerline
{\tfu Apartado de Correos 99, 03080 Alicante, Spain}

\centerline
{\tfu 2. Physikalisches Institut, Universitaet Bern}

\centerline
{\tfu Sidlerstrasse 5, 3012 Bern, Switzerland}

\vskip 0.3truecm

\centerline{\tfu 3. Observatori Astron\`omic,
Universitat de Val\ez ncia}

\centerline
{\tfu Av. Vicent Estell\es s s/n, Burjassot 46100,
Valencia, Spain}

\vskip 0.3truecm

\centerline
{\tfu 4. Departamento de Matem\`atica Aplicada y Estad\'{\i}stica}

\centerline
{\tfu Universidad Polit\'ecnica de Cartagena, Paseo Alfonso XIII,
30203 Cartagena (Murcia), Spain}

\vskip 1truecm

\centerline {\tfdue Accepted for publication in {\ifdue Icarus}}

\vfill\eject

\baselineskip 18 pt

\tfz

\centerline{\bfz Abstract.}

Dohnanyi's (1969) theory predicts that a collisional system
such as the asteroidal population of the main belt should rapidly
relax to a power-law stationary size distribution of the kind
$N(m)\propto m^{-\alpha }$, with $\alpha $ very close to $11/6$,
provided all the collisional response parameters are independent
on size.
The actual asteroid belt distribution at observable sizes, instead,
does not exhibit such a simple fractal size distribution.

We investigate in this work the possibility that the corresponding
cumulative distribution may be instead fairly
fitted by multifractal distributions.
This multifractal behavior, in contrast with the Dohnany fractal
distribution, is related to the release of his hypothesis of
self-similarity.

\vskip 2 truecm

\noindent {\bfu Keywords} Asteroids -- Collisional Physics --
Planetesimals

\vfill\eject \topskip 0 truemm \footline={\hss\tenrm\folio\hss}

\baselineskip 18 pt

\noindent {\bfdue 1. Introduction}

\tfu

\noindent A collisional system may be defined as a population of
bodies of different sizes, interacting with each other through
occasional collisional events, whose probabilities per unit time
are given functions of the size of the projectile and the target.
As a consequence of an impact, both bodies are converted into a
set of fragments whose size distribution depends on the relative
size of the colliding bodies, their relative impact velocity, and
a number of {\ifu collisional response} parameters determined by
the composition, mineralogy, and possibly also their previous
collisional history (Davis et al., 1992, 1993, Asphaug, 1999,
Durda and Flynn, 1999).  In planetary science, at least five examples
of such collisional systems are often discussed: accreting
planetesimals, planetary rings, the asteroid belt, the Trojan
swarms systems and the Edgeworth--Kuiper Belt.
To some extent, the case of the asteroid belt is the simplest one,
since collisional velocities are so high that impacts mainly give
raise to disruptive processes.

A useful mathematical model for the collisional evolution of the
asteroid belt was introduced by Dohnanyi (1969) and
independently by Hellyer (1970; for the sake of brevity, we
shall refer hereinafter to Dohnanyi's work only).  This model is
based on one crucial assumption: all the collisional response
parameters must be size-independent, implying that the transition
from cratering to fragmentation outcomes occurs for a fixed
projectile--to--target mass ratio, and no self-gravitational
reaccumulation of fragments is taken into account.

The most important result of Dohnanyi's work is that, since
collisional process gives raise to a cascade of fragments shifting
mass toward smaller and smaller sizes, a simple power-law mass
distribution is approached.   Dohnanyi's theory states that
the number of bodies $dN$ within the mass interval $(m, m+dm)$ is
proportional to $m^{-11/6} dm$, or---in the diameter interval $(D,
D+dD)$---to $D^{-7/2} dD$, with the proportionality coefficients
decreasing with time as the disruptive processes go on.  The
relaxation to this stationary mass distribution is fast: in the
asteroid belt it occurs over a time span much shorter than the age
of the solar system.
Other authors (Williams and Wetherhill, 1994; Paolicchi, 1994; and
Tanaka et al., 1996) confirmed these results and stated that the
stationary exponent of the resulting power-law distribution is
independent on the model of the collisional outcome of the
fragmentation process, as long as the model remains self-similar.
Moreover, they showed that the value of the exponent itself is
determined only by the mass--dependence of the collisional rate
(Tanaka et al., 1996). Again, Williams and Wetherill (1994) have
shown that the $-11/6$ exponent changes very little, less than
$10^{-4}$, when Dohnanyi's collisional physics assumptions are
varied in a substantial way, but again, if the whole process
remains self-similar. Let's stress that all along this paper we
talk of self-similarity in a physical sense, that is referred to
the physical characteristics of the bodies taking part to the
collisional evolution.

On the other hand, almost complete and reliable observational data
for asteroids of size larger than some 10 km have been collected
in the past from direct optical and radar observations. From the
analysis of these available data it is straightforward to see that
in the asteroid belt distribution such a predicted fractal pattern
does not show up at all. The observed size distribution---when
represented on a double logarithmic plot---``bends'' in between
the large mass end and the completeness size, instead of being
represented by a straight line, typical of a fractal behavior.
This observed pattern has sometimes been mimicked by using some
different power-law distributions matched at some transition size.

As a matter of fact, the basic assumptions of Dohnanyi's theory
are not fulfilled in real asteroid collisional systems.
Collisional response parameters are not size--independent at all
(at least at sizes greater than a few km), as will be explained in
Sec.3. This affects the resulting mass distribution in a
substantial way, (Fujiwara et al., 1989; Davis et al., 1989, 1994;
Campo Bagatin et al., 1994; Campo Bagatin, 1998).

\noindent {\bfu 2. Multifractal distributions}.

It is possible to model a fragmentation process by means of a
multifractal description. To illustrate how this can be done, we
shall use an analytical multifractal mass distribution introduced
in the analysis of turbulence by Benzi et al. (1984) (see also
Menevau \& Sreenivasan (1987), and Tel (1988)). Let us consider
the unit interval and subdivide it into $k$ equal pieces of size
$1/k$. Assign to each piece a probability, or measure,
$\{p_i\}_{i=1}^k$ such that $\sum_{i=1}^k p_i =1$. We continue
this process by subdividing each segment of length $1/k$ into $k$
new segments of length $1/k^2$; now the probabilities
$\{p_i\}_{i=1}^k$ are assigned to each segment following the same
deterministic pattern of the first state distribution. The measure
associated to each of the new intervals is just the product of
this new assigned value by the $p_i$ value of its parent interval.
The subdivision procedure is repeated again and again. After $L$
steps, we have $k^L$ intervals of length $1/k^L$, each with a
measure being the product of the last $p_i$ assigned to that
interval by all its ancestors. The distribution of the measure
becomes very inhomogeneous after a few steps of this construction.
This kind of process is named ``self-similar" in mathematics, as
it is generated always in the same way, but this should not be
confused with the physical definition of ``self-similarity" given
in Sec.1. The distribution itself is a distribution of mass (see
Falconer 1990) and illustrates an
inhomogeneous---multifractal---distribution on a compact
non--fractal support (the unit interval).

We can see in Fig. 1 how the concentration of mass (represented by
the height of the bars) varies widely from one region to another
within the unit interval, for the case $k=3$. (Note that if one of
the values of the probabilities is zero, for example $p_2=0$, and
the construction is continued indefinitely, the support of the
multifractal measure should be the triadic Cantor set, a fractal
with dimension $D=\log 2/\log 3$.)

In general, a given set is said to show a multifractal behaviour,
if the corresponding statistical analysis takes into account the
number of its elements as well as their spatial distribution. In
the case shown here the analysis is made only on the number of
elements---and the resulting distribution is multinomial---but we
shall rather keep the term because the analysis as a whole is
indeed a multifractal one.

On the other hand, many different physical phenomena related to
the natural processes of multiplicative cascades have been
explained in terms of fractal and multifractal distributions
(Falconer 1990, Meneveau and Sreenivasan, 1987, Martinez 1990,
Borgani 1993, Chiu and Hwa 1991, Takagi 1994). We introduce here a
multifractal technique for the analysis of the collisionally
evolved population of main belt asteroids as a natural tool to
treat this kind of physical systems.

\noindent {\bfu 3. Stationary distributions and self-similarity.}

If we think about the stationary distribution for any collisional
system in terms of fractal distributions, we see it can be
characterized by its fractal dimension $d_0$, that is by the
absolute value of the exponent of the distribution $N(\epsilon)
\approx \epsilon ^{-d_0}$, where $\epsilon = m/M$, $m$ is a
generic mass, $M$ is the whole mass--range considered and
$N(\epsilon)$ is the number of objects with mass $m$.  The
explicit relationship modelling such a distribution in a
collisional system---as expressed in terms of the number of
objects $dN$ in the mass interval $dm$---is of the kind
$dN(m,m+dm)\propto m^{-\alpha}dm$, with $\alpha=11/6$ under
Dohnanyi's hypotheses. Normalizing over $M$ we can identify $d_0$
with $\alpha$, that is the fractal exponent coincides with the
Dohnanyi stationary exponent.

\par

The only reliable data about the distribution of the number of
objects at a given size in the \mab are the ones corresponding to
the size distribution of \ast in the observable range, that is
from a few km--size to Ceres' 913 km. The distribution of masses
($M$) can be related to that of sizes (diameters, $D$) via the
obvious relationship $M={\pi \rho D^3 \over 6}$, assuming a
constant density $\rho$---likely between 1.5 and 3.5
g/cm$^3$---and spherical shapes.

By analyzing this distribution, it can be easily checked that the
observed size distribution is not at all a simple power-law
(fractal) distribution (see Fig. 2). The fact that \ast in this
size range do not follow the analytical distribution claimed by
Dohnanyi and others is not surprising to us. In fact, one of the
hypotheses that lead to the fractal stationary distribution is
released here: there is {\ifu no} self-similarity in the
collisional processes of asteroids of observable sizes due to two
main reasons.

{\ifu (a)} Fragmentation experiments and models suggest a set of
scaling--laws in order to accomplish for a realistic description
of the resistance of materials to high velocity fragmentation. A
typical physical magnitude that is useful in characterizing
fragmentation phenomena is the {\ifu impact strength}, defined as
the threshold energy density (energy per unit volume) necessary to
fragment a target in a way that the greatest fragment produced has
$50\%$ the mass of the target. Scaling laws take into account both
effects due to: {\ifu (i)} gravitational { \ifu
self--compression}, that is an increase of the strength ($S$) due
to the pressure exercised by the upper material layers upon the
deepest ones, increasing with the square of target size (Davis et
al., 1985): $S^´ =S_0+A(\rho D)^2$ ($S_0$ is the impact strength
measured in laboratory experiments, $A$ is a constant);{\ifu (ii)}
the so--called {\ifu strain--rate} effect, which is suggested by
the idea that the specific energy necessary for a breakup should
decrease as the size of the target increases (Farinella et al., 1982,
Housen et al. 1991, Housen and Holsapple 1999): $S\propto S^´
/D^{q}$ ($q$ ranges from $1/4$ to $1/2$).

\par

{\ifu (b)}  The {\ifu self-gravity} of celestial bodies, on the
other hand, varies as well at different sizes, and one of its
effects is that---apart from the self-compression quoted
above---it directly affects the escape velocity, especially in
targets of size greater than at least a few hundred meters (the
transition size is not univocally determined, as discussed in
Holsapple, 1994; Love and Ahrens, 1996; Melosh and Ryan, 1997;
Benz and Asphaug, 1999): then---for a fixed mass of projectile to
mass of target ratio ($M_p/M_t$)---the mass ratio of the escaping
fragments divided by the mass of the target is not independent on
$M_t$ itself. In other terms, the specific energy for disruption
Q$_D^*$ (that is the energy per unit mass necessary to disperse
``to infinity'' half of the mass of the target) is a function of
$M_t$. This effect, by itself, is a clear example of violation of
the self-similarity condition.
This significantly modifies the characteristics of the
multiplicative cascade, contrarily to what happens in systems in which reaccumulation
is non--existing or negligible. But reaccumulation is a common
collisional outcome for asteroids, the relative amount of
reaccumulated asteroids at observable sizes has been estimated to
range between 0.5 and 1 (Campo Bagatin et al., 2001).
Finally, a third potential source for non--self-similarity can be
deduced from the experimental results reported by many authors
(Davis and Ryan, 1994, Nakamura and Fujiwara, 1991, Giblin,
1998), indicating a shallow mass--velocity dependence in the
velocity distribution of the ejected fragments of a catastrophic
collision.

>From what shortly described here, it is obvious that the quoted
effects imply lack of self-similarity in the physics of
fragmentation at different sizes. The resulting stationary
distribution may  depart from the well-behaved fractal predicted
by the Dohnanyi theory and it may need to be represented by the
structured multifractal distributions proposed in this work.

\noindent {\bfu 4. Multifractal analysis}.

The goal of this research is to point out and show that
multifractal analysis may be a natural and useful tool to match
the observed stationary distributions of collisional systems when
self-similarity assumptions in the characteristics of the
collisional cascade are not expected to hold.

The multifractal analysis, and the corresponding fits to the
observed populations of the \mab  were obtained by means of the
following simple procedure. First, we set a size interval ranging
from around 30 km ($D_{\min}$) to about 400 km ($D_{\max}$) for
asteroid sizes, and we consider the corresponding cumulative
population of estimated masses in the asteroid belt ($M_{\min}$,
$M_{\max}$). We have chosen this range because it is the more
reliable as far as completeness of data is concerned. There are
only 3 asteroids larger than 400 km, but---due to their
size---they are not likely to have been the outcomes of the
collisional evolution process since the time it reached the
present stationary regime, some 4.5 Gyr ago. Quantitatively, it
can be easily checked from previous works on the collisional
evolution of the asteroid belt (Campo Bagatin, 1998) that
shattering and dispersal of Ceres--size objects, the ones that
could produce any single 400--500 km body present in the actual
asteroid distribution, is expected approximately once every 4--5
Gys. Since this means that only one of the asteroids larger than
400 km could have been contributing to the collisional cascade,
our assumption of neglecting the largest bodies in the collisional
process seems reasonable, thence they have not been included into
our statistical study.

Then we look for a multifractal distribution described by a set of
multinomial probabilities ($p_i=p_1,\ p_2,\ ,\dots,\ p_k$), such
that any given term of the kind $p_1^{r_1}p_2^{r_2}\cdot \cdot
\cdot p_k^{r_k}$ corresponds to some mass in the actual
distribution, in such a way that the cumulative multiplicity of
any given term is equal to the cumulative number of bodies of the
corresponding mass. The choice of comparing cumulative
distributions rather than differential ones is due to avoid
comparing artificial subdivisions of the asteroidal mass range in
discrete arbitrary bins, with the multifractal mass distribution.
Multinomial terms are transformed to masses by scaling their range
to the corresponding given mass range ($M_{\min}$, $M_{\max}$),
chosen accordingly to the size range from about 30 km to about 400
km, as explained above.

In order to derive the probabilities $p_i$ we therefore perform a
systematic search for sets of probabilities matching the observed
distribution, according to an automatic least-squared method, and
when we find sets that fulfill the matching conditions, we finally
transform back masses to diameters in order to get the plots in
Fig.2, in which the observed size distribution and the obtained
fits are shown.

At the end of the systematic search we found that for some sets of
$k,\ L$, and $p_i\ (i=1,k)$ (shown in Table 1), the matching
between the observed distribution and the multifractal ones is
fairly good. We found sets of probabilities fitting a large number
of multifractal terms and asteroid masses (sizes), corresponding
to low dimensions $k$, that is to small number of parameters
represented by the probabilities $p_i$. That happened for $k=2$
and $k=3$, for which the observed distribution could be best
fitted with up to 12 multifractal terms, depending on the
different values of the multiplicative level L.

The distributions corresponding to the best matching sets can be
seen in Fig. 2A to 2D. The deviations of any cumulative
multiplicity of each multifractal term from its corresponding
cumulative asteroid number is normally below 20$\%$ in any of the
$k=2$ and $k=3$ cases.

A source of uncertainty in the translation from multifractal terms
to masses to diameters of asteroids is the one on the value of the
density, that we assume to be 2.5 g/cm$^3$. On the other hand, no
error bars are shown on plots for the number of observed asteroid
at any given size, but it should be pointed out that at least two
possible sources for uncertainty in the observed asteroidal size
distribution (and the related mass distribution) may affect the
present analysis and the sets of probabilities:

i) the estimation of the sizes of asteroids, obtained from
observed magnitudes and assumed surface albedos;

ii) the actual cumulative number of asteroids, that should be
anyway rather reliable for sizes larger than 50 km, for fixed
estimations of albedo and density.

\noindent {\bfu 5. Results and conclusions}.

A set of reliable fits of the proposed statistical distributions
to the observable data for the cumulative size distribution of
main belt asteroids larger than 20 km is shown in Fig. 2A through
2D. We have found out a set of different multifractal
distributions that may match fairly well the current cumulative
distribution of the asteroid belt at observable sizes. This is
interesting both on an operative and a physical ground. In fact,
if the technique introduced here is used for other sets of
collisionally evolved systems, it could provide a useful tool to
match such distributions in a reliable way, with no need of
guessing different power-law ranges matched somehow to
approximately reproduce the data, as it has been done up to now:
all we need is the mass distribution, or the size distribution
with given assumptions about the shape and density of the bodies.

>From a physical point of view, the fact that it is possible to fit
the population of the main asteroid belt---or of any other
nonself-similar collisonal system---by means of a multifractal
distribution, introduces the fresh idea that physical
inhomogeneities in fragmentation processes produce multifractal
structures in this kind of systems. That also shows clearly that
in these cases the Dohnanyi's hypotheses cannot be assumed and his
conclusions cannot be generally applied.

But, what is the exact physical meaning of the found probabilities
$p_i$s? It is not obvious how they may be related to collisional
physics and/or to the conditions for the evolution. We suggest
that such a correlation is embedded in the collisional cascade
process described by multifractal distributions, as it happens to
be for other multiplicative systems. We are investigating this
major issue, in order to relate quantitatively the multifractal
parameters to the physical characteristics of collisional
evolution, but we prefer, at this stage, to limit the present
interpretation to a phenomenological analysis, and to postpone the
answer to this main question.

As far as the asteroid belt is concerned, we observe that at sizes
smaller than a few tens of km---that is outside the range of the
present work---extrapolations of recent data (Jedicke and Metcalfe,
1998) show a quasi-fractal behavior. This is not in contrast with
what we have found in the present work. In fact, in the size range
in which self-similarity does not hold, a depart from a simple
fractal behavior due to nonself-similarity is expected and
observed, leading to the multifractal distribution that we have
introduced in this work. On the other hand, as soon as
self--similarity begins to show up---like it seems to happen for
smaller size bodies, for which self-gravitational effects have no
main influence---a fractal behaviour becomes possible, as Dohnany
and the other quoted authors have shown in the past. Campo Bagatin
et al. (1994) and Campo Bagatin (1998) also showed that the release of one of the
simplifying Dohnanyi's hypothesis---that is the absence of a lower
cutoff in the mass distribution of asteroids---lead to a ``wavy''
pattern in the finally evolved distribution. Here we have shown a
consequence of releasing another simplifying hypothesis, that of
self-similarity. The two effects could in principle combine in
shaping the actual mass distribution of asteroids in the main
belt, making the whole picture more difficult to put into a unique
simple scheme.

\vfill \eject

\centerline {\bfu References}

\vskip 2 mm

{\frenchspacing \parindent=0mm \parskip=2mm

\baselineskip 18pt

Asphaug, E. 1999, Survival of the weakest, {\ifu Nature}, {\bfu 402}, 127--128.

Benz, W., E.Asphaug 1999, Catastrophic disruptions revisited, {\it
Icarus \bfu 142}, 5--20.

Benzi, R., G. Paladin, G. Parisi, A. Vulpiani 1984, On the multifractal
nature of fully developed turbulence and chaotic systems, {\it J.Phys.A} {\bfu 17} 3521--3531.

Borgani S., G.Maurante, A. Provenzale, R.Valdarnini 1993, Multifractal
analysis of the Galaxy Distribution: Reliability of Results from Finite Data Sets, {\it Physical Review E}
{\bfu 47}  3879--3888.

Campo Bagatin, A., P. Farinella, and J--M. Petit 1994, Fragment
ejection velocities and the collisional evolution of asteroids, {\it
Planet. Space Sci.}, {\bfu 42}, 1099--1107.

Campo Bagatin, A. 1998, Collisional evolution of asteroidal systems, Ph.D.
Thesis {\ifu Servicio de Publicaciones Universidad de Valencia, Valencia, Spain}.

Campo Bagatin, A., J.-M. Petit, P. Farinella 2001, How many rubble piles are in the asteroid belt?, {\it Icarus} {\bfu 149}, 198--209.

Chiu C.B., R.C. Hwa 1991, Multifractal structure of multifractal production in
branching models,  {\it Physical Review D}
{\bfu 43} 100--103.

Davis, D.R., C.R. Chapman, S.J. Weidenschilling, and R. Greenberg 1985,
Collisinal history of asteroids: Evidence from Vesta and the Hirayama families,
{\it Icarus}, {\bfu 62}, 30--53. 

Davis, D.R., P. Farinella, P. Paolicchi, S.J. Weidenschilling, and
R.P. Binzel 1989, Asteroid collisional history: Effects on sizes
and spins,  in {\it Asteroids II}, eds. R.P. Binzel, T. Gehrels
and M.S. Matthews (Univ. of Arizona Press), pp. 805--826.

Davis, D.R., P.Farinella, P. Paolicchi, A. Campo Bagatin, A.
Cellino, and V. Zappal\`a 1992, Collisional effect of tiny
asteroids on their larger siblings, {\it Bull. Amer. Astr. Soc.}
{\bfu 24}, 961.

Davis D.R., P.Farinella, P.Paolicchi, A. Campo Bagatin, A. Cellino.
and V. Zappal\`a 1993, Deviations from the straight line: Bumps
(and grinds) in the collisionally evolved size distribution of
asteroids, {\it Lunar Planet. Sci. Conf. XXIV}, 377--378.

Davis, D.R., E.V. Ryan, and P. Farinella 1994, Asteroid
collisional evolution: Results from current scaling algorithms.
{\it Planet.  Space Sci.}, {\bfu 42}, 599--610.

Donhanyi J.W. 1969, Collisional model of asteroids and their
debris, {\it J. Geophys. Res.} {\bfu 74}, 2531--2554.

Durda, D.D., G.J.Flynn 1999, Experimental study of the impact
disruption of a porous, inhomogeneous target, {\ifu Icarus}, {\bfu
142}, 46--55.

Falconer, K.J. 1990. Fractal geometry. Mathematical foundations
and applications, {\ifu John Wiley \& Sons Eds.}.

Farinella, P., P. Paolicchi, V. Zappal\`a 1982, The asteroids as
outcomes of catastrophic collisions, {\ifu Icarus}, {\bfu 52},
409--433.

Fujiwara, A., P. Cerroni, D.R. Davis, E. Ryan, M. Di Martino, K.
Holsapple, and K. Housen 1989, Experiments and scaling laws on
catastrophic collisions, in {\it Asteroids II}, Eds.  R.P. Binzel,
T. Gehrels and M.S. Matthews (Univ. of Arizona Press), pp.
240--265.

Giblin, I. 1998, New data on the velocity--mass relation in
catastrophic disruption, {\ifu Planet. Space Sci.} {\bfu 46},
356-364.

Hellyer, B. 1970, The fragmentation of the asteroids, {\it Mon.
Not. R. astr. Soc.} {\bfu 148}, 383--390.

Holsapple, K.A. 1994, Catastrophic disruptions and cratering of
the Solar System: A review and new results, {\it Planet. Space
Sci.}, {\bfu 42}, 1067--1078.

Housen, K.R., Holsapple, K.A. 1999, Scale effects in strength
dominated collisions of rocky asteroids, {\ifu Icarus}, {\bfu
142}, 21--33.

Housen, K.R., R.M.Schmidt, and K.A.Holsapple 1991. Laboratory
simulations of large scale fragmentation events, {\ifu Icarus},
{\bfu 94}, 180--190.

Jedicke R., and T.S. Metcalfe 1998, The orbital and absolute magnitude
distribution of main belt asteroids, {\it Icarus} {\bf 131}, 145--260.

Love, S.G., and T.J. Ahrens 1996, Catastrophic impacts on gravity
dominated asteroids, {\ifu Icarus}, {\bfu 124}, 141--155.

Mart\'{\i}nez V.J. 1990, Fractals and multifractals in the
description of the cosmic structure, {\ifu Vistas in Astronomy},
{\bfu 33}, 337--356.

Melosh H.J., and E.V.Ryan 1997, Asteroids: Shattered not
dispersed. {\ifu Icarus}, {\bfu 129}, 562--564.

Meneveau, C. and K.R. Sreenivasan 1987, Simple multifractal
cascade model for fully developed turbulence, {\it
Phys.Rev.Lett.}, 1424--1427.

Nakamura, A., Fujiwara, A. 1991, Velocity distribution of
fragments formed in a simulated collisional disruption, {\ifu
Icarus}, {\bfu 92}, 132--146.

Paolicchi, P. 1994, Rushing to equilibrium: A simple model for the
collisional evolution of asteroids, {\it Planet. Space Sci., {\bfu
42}, 1093--1097}.

Takagi, F. 1994, Multifractal structure of multiplicity
distributions in particle collisions at high energies, {\it
Physical Review Letters} {\bfu 72} 32--35.

Tanaka, H., S. Inaba, and K. Nakazawa 1996, Steady--state size
distribution for the self--similar collision cascade, {\ifu
Icarus}, {\bfu 123}, 450--455.

Tel, T. 1988, Fractals, multifractals and thermodynamics, {\ifu Z.
Naturforsch.}, {\bfu 42a}, 1154--1174.

Williams, D.R., and G.W. Wetherill 1994, Size distribution of
collisionally evolved asteroid populations: Analytical solution
for self--similar collision cascades, {\it Icarus} {\bfu 107},
117--128.

\vfill\eject

\baselineskip 18pt

\tfz

\medskip

\vbox{\tabskip=0pt \offinterlineskip

\def\tablerule{\noalign{\hrule}}

\halign to219pt{\strut#& \vrule#\tabskip=1em plus2em& \hfil#& \vrule#&

\hfil#\hfil& \vrule#&

\hfil#\hfil& \vrule#&

\hfil#\hfil& \vrule#&

\hfil#\hfil& \vrule#&

\hfil#\hfil& \vrule#\tabskip=0pt\cr\tablerule

&&\omit\hidewidth  Case    \hidewidth&&
\omit\hidewidth  A   \hidewidth&&
\omit\hidewidth  B   \hidewidth&&
\omit\hidewidth  C   \hidewidth&&
\omit\hidewidth  D   \hidewidth&\cr\tablerule
&& k && 2 && 2 && 3 && 3 &\cr\tablerule
&& L &&  11 && 12 && 9 && 11 &\cr\tablerule
&& $p_1$ && 0.818 && 0.833 && 0.737 && 0.789 &\cr\tablerule
&& $p_2$ && 0.182 && 0.167 && 0.168 && 0.178 &\cr\tablerule
&& $p_3$ && -- && -- && 0.095 && 0.033 &\cr
\tablerule
\noalign{\medskip}\hfil\cr}}

\tfu

{\bfu Table I}  The 4 sets of $k, L,\ p_1,\ p_2,\ p_3$ leading
respectively to the fits
represented in Fig. 4A, B, C, D.


\baselineskip 18pt

\vskip 5mm

\noindent {\bfu Figure captions}

\vskip 4 mm

{\frenchspacing \parindent=0mm \parskip=3mm

{\bf Figure 1.} A few steps of the construction of the
self-similar multifractal measure described in the text by means
of a multiplicative cascade. The values of the initial
probabilities are $p_1=0.3534$, $p_2=0.4363$ and $p_3=0.2103$. The support of
the measure is the unit interval. At each stage of the
construction, the measure associated to each interval is
represented by the height of the rectangular box. We
show steps 1, 2, 4, and 8, from top to bottom and 
left to right. After several
steps, the distribution of the measure becomes rather
inhomogeneous, a characteristic pattern of multifractal
distributions.

{\bf Figure 2.} The four best fits to the main asteroid belt
cumulative population of asteroids, using the multifractal
technique explained in the text. The dashed lines show the
observed distribution and the crosses represent 
the multifractal fits.

Fig. 2A to 2D correspond respectively to cases A to D shown in Table 1.

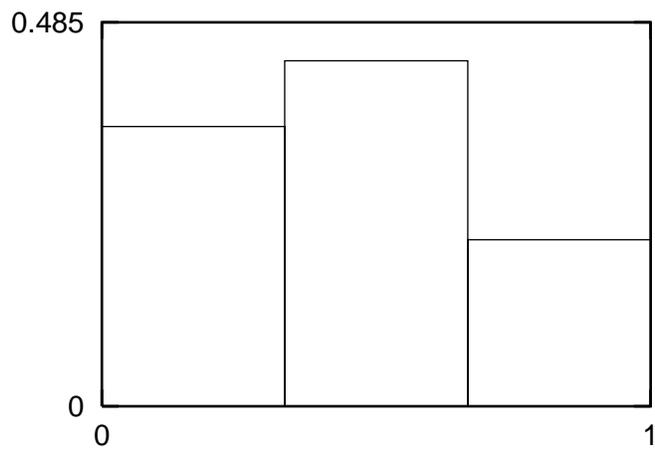
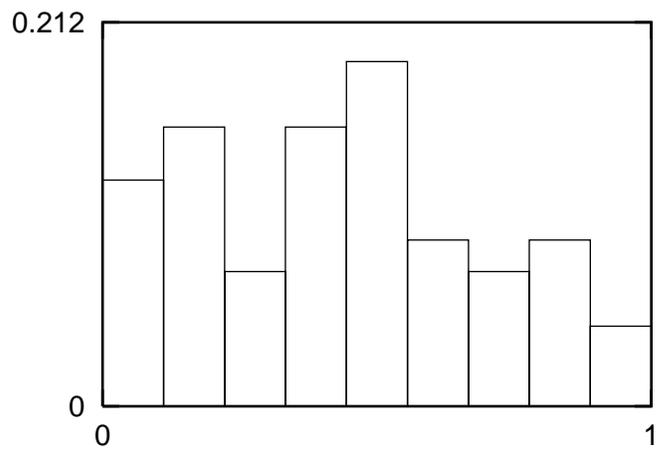
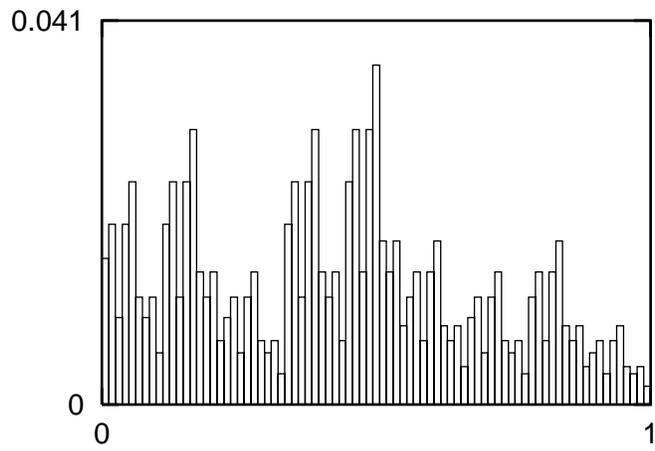
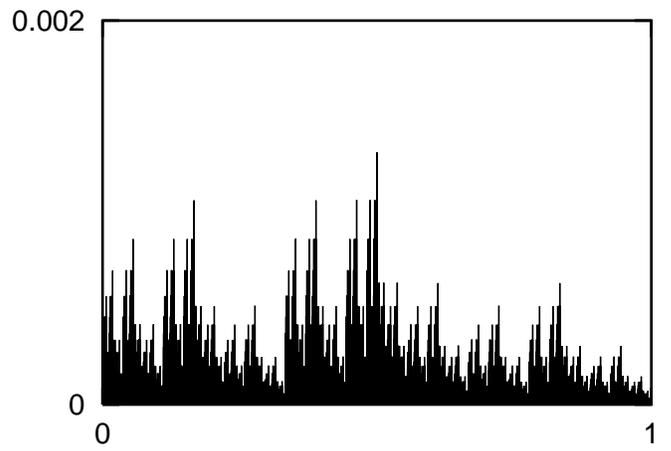
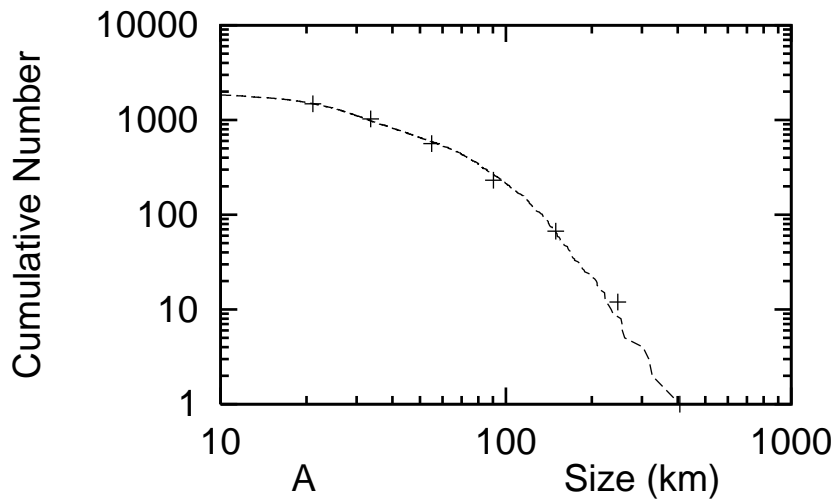
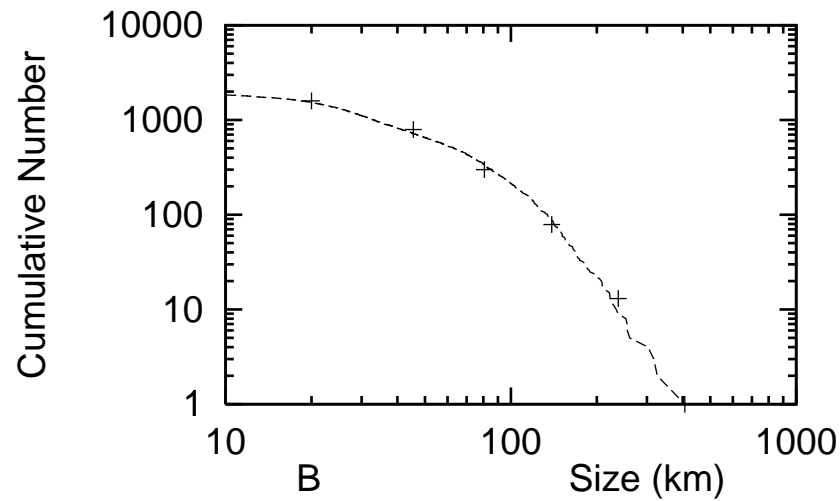
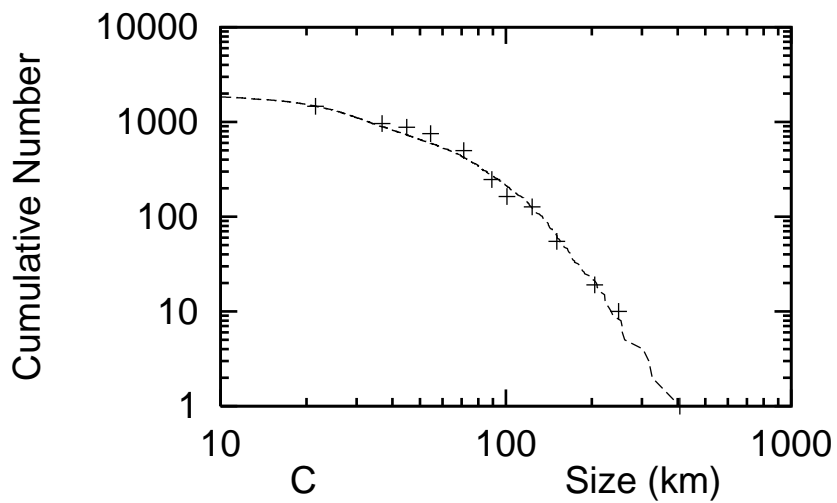
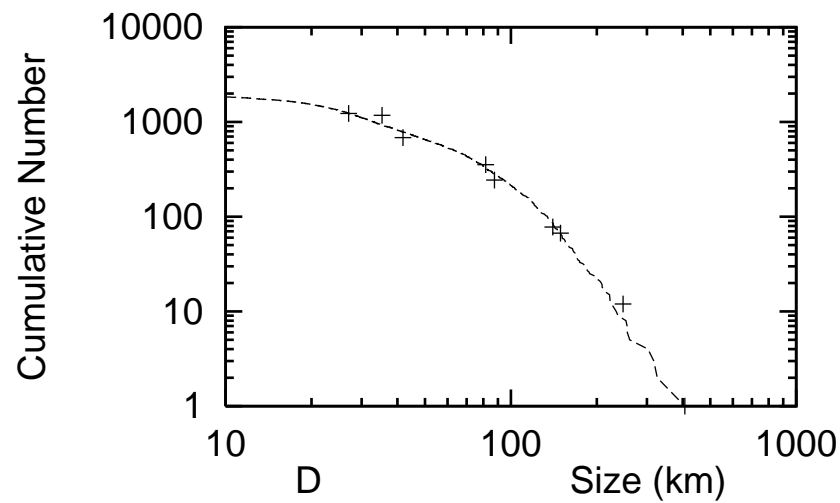
\bye